# Transparent Metasurfaces Counteracting Fogging by Harnessing Sunlight

*Christopher Walker†, Efstratios Mitridis†, Thomas Kreiner, Hadi Eghlidi, Thomas M. Schutzius\*, Dimos Poulikakos\**

Laboratory of Thermodynamics in Emerging Technologies, Department of Mechanical and Process Engineering, ETHZ Zurich, Sonneggstrasse 3, CH-8092 Zurich, Switzerland

† Denotes equal contribution

\* To whom correspondence should be addressed

Prof. Dimos Poulikakos
ETH Zurich
Laboratory of Thermodynamics in Emerging Technologies
Sonneggstrasse 3, ML J 36
CH-8092 Zurich
Switzerland
Phone: +41 44 632 27 38
Fax: +41 44 632 11 76
dpoulikakos@ethz.ch

Dr. Thomas M. Schutzius
ETH Zurich
Laboratory of Thermodynamics in Emerging Technologies
Sonneggstrasse 3, ML J 27.2
CH-8092 Zurich
Switzerland
Phone: +41 44 632 46 04
thomshu@ethz.ch

# Abstract

Surface fogging is a common phenomenon that can have significant and detrimental effects on surface transparency and visibility. It affects the performance in a wide range of applications including windows, windshields, electronic displays, cameras, mirrors, and eyewear. A host of ongoing research is aimed at combating this problem by understanding and developing stable and effective anti-fogging coatings that are capable of handling a wide range of environmental challenges "passively" without consumption of electrical energy. Here we introduce an alternative approach employing sunlight to go beyond state-of-the-art techniques—such as superhydrophilic and superhydrophobic coatings— by rationally engineering solar absorbing metasurfaces that maintain transparency, while upon illumination, induce localized heating to significantly delay the onset of surface fogging or decrease defogging time. For the same environmental conditions, we demonstrate that our metasurfaces are able to reduce defogging time by up to four-fold and under supersaturated conditions inhibit the nucleation of condensate outperforming conventional state-of-the-art approaches in terms of visibility retention. Our research illustrates a durable and environmentally sustainable approach to passive anti-fogging and defogging for transparent surfaces. This work opens up the opportunity for large-scale manufacturing that can be applied to a range of materials, including polymers and other flexible substrates.

The loss of visibility due to surface fogging is a common phenomenon, which presents itself in a variety of daily situations, affecting both transparent and reflective surfaces, such as windows, windshields, electronic displays, cameras, mirrors and eyewear including eyeglasses, safety glasses, ski and scuba goggles, and face masks.[1] This visibility loss is a result of micro-sized water droplets that nucleate and grow on the surface due to either a sudden increase in relative humidity or a sudden drop in surface temperature. These water droplets disperse and reflect the incident light, severely affecting optical clarity.

A number of concepts have been studied and implemented to suppress this undesirable phenomenon. Temporary passive solutions include surfactants or superhydrophilic coatings, both of which increase the relative energy of the surface and shift the equilibrium from a plethora of tiny individual droplets to a very thin and continuous layer.[2,3] In contrast to individual droplets, this thin and continuous water layer does not scatter the incident light, and therefore does not interfere as much with the transparent clarity of the surface. Although improvements have been made to render superhydrophilic coatings more robust,[4] these coatings are more susceptible to contamination due to their increased surface energy[5] and are prone to water inundation, which can reduce visibility. Research aimed at addressing this concern has concentrated on developing zwitter-wettable surfaces, which exhibit either hydrophobic or hydrophilic properties dependent on the water droplet residence time on the surface.[6]

Permanent passive solutions typically take the approach of implementing superhydrophobic self-cleaning surfaces to remove condensed water droplets from the surface. Intricately-engineered superhydrophobicity is here necessary to alleviate nanoscale condensate that nucleates within the textures and destroys the stability of the Cassie-Baxter state and therefore their hydrophobicity property in environments of high relative humidity.[7,8] Both growing condensate droplets as well as new macroscale droplets that come into contact with the surface will be forced into the Wenzel state, producing pinned droplets, destroying the desirable superhydrophobic characteristics,[9] and thus rendering the surface even more adhesive than an otherwise flat surface of same chemical composition.[10] There have also been approaches to mitigate the loss of hydrophobicity resulting from surface condensate, and therefore render such sustainably hydrophobic surfaces effective for anti-fogging. Many of these approaches, inspired by surface features observed on insect eyes and plant leaves, include creating textured surfaces whose size and rationalized geometry allows condensate to grow in the Cassie-Baxter state.[11–14] Because the Cassie-Baxter state preserves the droplet mobility, these droplets can be easily removed by small external forces, such as wind or gravity. An additional and extensively investigated approach is to use the excess energy of coalescing droplets, resulting from the reduction in free surface area, to propel themselves off the surface.[15–19] While both of these approaches enhance the anti-fogging performance, they do not reduce the likelihood of the formation of the undesired condensate in the first place, which, in particular over time, can have a detrimental effect for certain applications.

To address this challenging problem, we propose a different approach. As in recent pathbreaking studies revolving around harnessing light and sunlight for new applications,[20–23] including catalysis, materials synthesis, and desalination,[24–27] we utilize a sustainable solution by means of rationally designed transparent natural sunlight absorbing surfaces,[28,29] hereafter termed as metasurfaces, designed to inhibit the heterogeneous nucleation of condensate and also exhibit superior anti-fogging performance, while maintaining optical transparency. We begin by showing qualitatively and quantitatively the shortcomings of three different kinds of wetting behavior on a plain glass substrate, ranging from superhydrophilic to hydrophobic. Additionally, we show that the light-absorbing metasurfaces of the present work enhance evaporation rates of already accumulated condensate. To this end, we analyze, both experimentally and theoretically, the condensation and evaporation behavior of a single droplet on control surfaces and our metasurfaces. This fundamental analysis and discussion provide the motivation to subsequently consider condensation and evaporation of multiple droplets. For our metasurfaces, we observe significant reduction in condensate nucleation rate and an increased evaporation rate. We believe that this research, as a stand-alone technology, as well as in combination with the previously described state-of-art research, will lead to more robust and enhanced passive anti-fogging and defogging surfaces.

## Results and Discussion

In order to understand and place into perspective the inherently intertwined phenomena of visibility diminution and surface fogging for transparent and reflective media, we began by investigating the behavior of superhydrophilic (the most common state-of-the-art technique to retain visibility), hydrophobic glass, and untreated glass with intermediate wetting behavior ($\theta \approx 90°$), under exposure to a supersaturated water vapor environment. Throughout this study we compare these glass surfaces to our sunlight absorbing metasurfaces, made by sputter deposition. This is a flexible coating technique that has been demonstrated in a roll-to-roll multilayer process for depositing transparent conducting oxides on windows at large-area,[30] while its applications also extend to highly-controllable metallization[31] and deposition of tiny amounts of noble metals with subsequent cost reduction.[32] By using this coating technique, we fabricated a metal-dielectric nanocomposite of gold nanoparticles embedded in titania on glass substrates (for more information on the fabrication and characterization of the metasurface, see Methods and Supporting Information, Figure S1 and Figure S4). Figure 1a–c shows superhydrophilic, untreated, and hydrophobic glass samples, respectively (see Methods for further details on sample preparation and experimental procedure, and Supporting Information, section "Advancing and receding water contact angles of the tested surfaces" and Figure S1, for contact angle measurements on all samples used in our study). The samples were placed on a background consisting of an array of dots, before ($t$ = 0 s) and after exposure for 10 min ($t$ = 600 s) to an environment of supersaturated water vapor (supersaturation ($p/p_\text{L}$) equal to 1.3). We used the array of dots as a reference to both qualitatively and quantitatively observe the clarity ($\bar{C}$) and distortion ($\bar{\delta}$) caused by condensate on the samples upon exposure to supersaturated vapor conditions.

Figure 1a compares a superhydrophilic glass surface before being exposed to supersaturated vapor ($t = 0$ s) and after exposure for 10 min ($t = 600$ s) (see Methods, section "Setup and experimental protocol" for further details). Comparing this to the same conditions for untreated and hydrophobic samples in Figure 1b–c, respectively, we clearly observe that the hydrophilic glass retains similar clarity as before exposure to supersaturated conditions, while the array of dots under the untreated and hydrophobic substrates suffer from significant blurriness due to the microdroplets scattering incident light. This supports the motivation for using superhydrophilic surfaces where maintaining transparency is important even when exposed to supersaturated vapor environments. Although the superhydrophilic glass sample retains the highest clarity among the three types of glass, it appears to suffer from distortion, apparent by the asymmetry of the array of dots at $t = 600$ s. In contrast, this asymmetry is not observable when inspecting the untreated and hydrophobic samples in Figure 1b–c.

To quantify our visual observations, we determined the parameters $\bar{C}$ and $\bar{\delta}$ (see also illustration in Figure 1d). We quantified $\bar{C}$ by measuring the mean intensity of each dot ($N = 25$ dots in total) at $t = 0$ s and $t = 600$ s and normalizing it to the respective mean background intensity ($\bar{I}_{b,0}$ and $\bar{I}_{b,t}$). We measured the intensities by taking the mean grayscale value of the pixels making up each dot and the background respectively ($\bar{I}_t$ and $\bar{I}_0$). We calculated $\bar{C}$ per each dot at time $t$ by:

$$\bar{C} = 1 - \log_{10}[(\bar{I}_t/\bar{I}_{b,t})/(\bar{I}_0/\bar{I}_{b,0})] \tag{1}$$

We quantified $\bar{\delta}$ by measuring the position of each dot, ($x_{i,0}, y_{i,0}$), at $t = 0$ s and ($x_{i,t}, y_{i,t}$), at $t = 600$ s, and using this to calculate the distance that each of the dots moves. We normalized this distance with the mean diameter of the dots ($d$), resulting in:

$$\bar{\delta} = \overline{\Delta s}/d = \frac{1}{d}\sum_{i=1}^{N}\sqrt{(x_{i,t}-x_{i,0})^2 + (y_{i,t}-y_{i,0})^2} \tag{2}$$

$\overline{C}$ and $\overline{\delta}$ are reported in Figure 1e and Figure 1f, respectively, as box plots, where the values of each box and whisker are made up of six experiments ($n=6$) and $N=25$ dots per experiment. As expected, Figure 1e confirms that the clarity of the untreated and hydrophobic glass surfaces is significantly reduced when compared to that of the superhydrophilic glass. We attribute the slightly better performance of the hydrophobic glass compared to the untreated glass to the reduction in nucleation density of the hydrophobic glass and therefore the reduction in the number of drops that scatter the light (see Supporting Information, sections "Theoretical effect of surface wettability on fogging resistance", Figure S2, and "Observed effect of surface wettability on fogging resistance", Figure S3). Furthermore, Figure 1f confirms that the distortion in the case of the superhydrophilic sample is considerable, especially when compared to the untreated and hydrophobic samples. An ideal hydrophilic surface should produce a perfectly continuous water film across the entire sample, however this is not what we observed. We attribute this to the high surface energy of the treated glass, making it more susceptible to contamination[33] that creates pinning points, acts to break up the continuous film, and results in the observable distortion. Even if the surface remained completely clean, additional pinning points from the edges would also create a curved water interface at the edges, and also result in distortion. In order to quantify and compare the overall visibility retention of the three surfaces (superhydrophilic, untreated and hydrophobic glass), we introduced a performance factor, $P=(1-\overline{\delta})\overline{C}$, that combines the values of both clarity and distortion. The higher the performance factor the better the visibility retention through the sample. We furthered our comparison to also consider the performance of our metasurface, assuming it remains condensate free (which we demonstate later). Figure 1g shows that the untreated glass is the worst performer, followed by the hydrophobic surface and the superhydrophilic surface. Finally, the metasurface achieves maximum performance (100%). It is important to understand that $\overline{C}$ does not take total light transmittance through the surface into consideration, therefore although our metasurface has a visible light transmittance of 36% compared to the glass transmittance of $\approx 87\%$, this does not affect $P$. We chose not to incorporate transmittance into $P$ because for many applications, a reduction in visible light transmittance is a desirable property to optimize user visibility comfort and performance. Applications such as car windows, building windows, and sun-protective eyewear utilize transmittance ranging from 15-50%,[34,35] 2-70%,[36] and 15-25%,[37,38] respectively. Advantageously, we designed the metasurfaces studied in this work to harvest both visible and near infrared light in order to obtain the most efficient anti-fogging performance. Important here is that although the metasurface reduces the sample transmittance, the maximum clarity and lowest distortion are achieved, due to the deeply sub-wavelength gold nanoparticles used to create the metasurface.

In order to demonstrate the design of our metasurfaces, we proceeded with characterizing their topography in the microscale and nanoscale. Figure 2a shows a schematic illustration of the different layers constituting the metasurfaces. In principle, they consist of repeated ultrathin bilayers of Au nanoparticles and a $TiO_2$ nanolayer. There is also an additional protective $TiO_2$ top-layer. Our metasurfaces have $k+1=4$ bilayers in total and a total thickness of $\approx 60$ nm. The cross-sectional composition of the metasurfaces is demonstrated in the scanning electron micrograph, which reveals that the layers of Au nanoparticles are embedded in and spaced apart by the $TiO_2$ layers. For improving image quality and emphasizing on the flexibility of the multilayer design, here we show the micrograph of a metasurface consisting of 22 bilayers. In Figure 2a, bottom-row, the atomic force micrograph of our metasurface indicates that the surface is in fact very smooth, with RMS roughness <1 nm. We then compared the performance of our metasurface to a glass substrate with a commercially available tinted laminate having similar level of transmittance in the visible spectrum. Figure 2b–c show the reflectance ($\mathcal{R}$), transmittance ($\mathcal{T}$), and calculated absorbance ($\mathcal{A} = 1 - \mathcal{T} - \mathcal{R}$) of both samples in the 400–1650 nm wavelength range. The visible light transmittance (400–750 nm) for the metasurface and tinted laminate was 36% and 33%, respectively. Figure 2b shows that the metasurface has a relatively constant broadband absorption in the measured wavelengths. On the other hand, Figure 2c shows that the commercially available tinted laminate has high absorption in the visible wavelengths (400–750 nm); however, in the near infrared range (750–1650 nm) its absorption drastically decreases. The broadband absorption of the metasurface is advantageous because it allows for increased absorbance of sunlight in the near infrared and therefore a higher temperature response to inhibit fogging or encourage evaporation (see Figure S5 for the spectra of the control sample). Figure 2d shows a schematic of the experimental setup and the light source spectrum that we used to illuminate the samples and measure their temperature. Figure 2e shows a plot of sample temperature above ambient, $\Delta T = \overline{T}_s(t) - \overline{T}_{s,0}$, vs time, $t$, upon illumination, for the metasurface and the tinted commercially available laminate, where $\overline{T}_s(t)$ and $\overline{T}_{s,0}$ are the average sample temperatures at time $t$ and $t = 0$ respectively. The steady-state value of $\Delta T$ for the metasurface was 130% higher than that of the commercially available tinted laminate (see also Supporting Information, section "Infrared thermography and temperature response measurements" for more information on the temperature response of the samples due to light-induced heating).

In order for the metasurfaces to achieve best performance in a supersaturated vapor environment, they must inhibit the formation of condensation on the surface. To this end, we studied the efficacy of our metasurfaces to use naturally occurring sunlight to generate plasmonic surface heating in order to aid in defogging and anti-fogging of surfaces exposed to supersaturated vapor environments, to maintain $\overline{C}$ and limit $\overline{\delta}$. Here we compared our metasurface to glass (hereafter referred to as control) for defogging and anti-fogging applications. Figure 3a illustrates the setup that we employed. It consisted of a sun-mimicking light source and lens to focus the light onto the sample plane (the metasurface was facing upwards) with a spot diameter of 7 mm that fully illuminated our 5 x 5 mm samples at a power density of 4 suns. We mounted two cameras to extract data in order to quantify the experiments; a CMOS angled side view camera to observe the droplet's contact angle and a top view infrared (IR) camera to construct the temperature profile of the sample and droplet. It is well-known that the contact angle progression of an evaporating droplet can have a significant effect on the evaporation time,[39–41] therefore in order to isolate the effect of the illuminated metasurface we treated both surfaces with a vapor-deposited trichlorovinylsilane (TCVS). This resulted in a consistent receding contact angle of approximately 50°, allowing consistent measurements between surfaces, beginning when the droplets reached the receding angle and then continued evaporation in the constant contact angle mode. Figure 3b quantifies the contact angle progression of the evaporating droplets as a function of time for both the control (red curves) and metasurface (blue curves) samples. We observed that the receding angle for both samples is the same and the constant contact angle mode becomes the mode of evaporation for the duration of the experiments. By establishing the same contact angle progression during evaporation for both surfaces, we show in Figure 3c that the metasurface increases the droplet evaporation rate by approximately four-fold for a droplet that is of the same order of magnitude compared to the sample size. This is caused by the considerable light absorption and heating of our metasurface, which is observable by examining Figure 3d and Figure 3e (see Supporting Information, section "Infrared thermography and temperature response measurements" and Figure S4 for more information on the temperature response of the samples due to light-induced heating). Here we show synchronized side-view (visible) and top-view (IR) image sequences of illuminated droplets on a control and metasurface sample, respectively, before and for the duration of evaporation. The leftmost IR images ($t < 0$ s) show the surfaces at steady state under illumination before a droplet is placed on the surface. While the control sample in Figure 3d remains very close to room temperature, the metasurface takes advantage of light induced localized surface plasmon oscillations on the nanometer-scale gold particles, resulting in heating of the surface by $>10$ °C. When placing a droplet on the surface, we observed a reduction in the temperature of the surface, shown in the IR images ($t = 0$ s) of Figure 3d and 2e. This can be explained by the effect of evaporative cooling, which causes the droplet, and subsequently the surface it is in contact with, to cool due to energy required by water molecules to overcome the latent heat of evaporation and change from liquid into vapor. Advantageous to the evaporation, the metasurface is able to transfer more heat into the droplet, while the droplet maintains a temperature lower than the surface temperature. Near the end of evaporation, the metasurface returns to its steady state temperature as in the case without the droplet, and so does the control, observable at $t = 200$ s and $800$ s in the IR images of Figure 3d and Figure 3e, respectively.

Using the ability of our metasurfaces to effectively harness sunlight energy over a broad spectrum and turn it into heat, we showed that we could reduce the drying time of single droplets. We proceeded to focus on the drying of fogged surfaces, as such an investigation is more relevant for surface defogging and anti-fogging. One would expect the enhanced evaporation rate in the case of the metasurface to also be present for multiple droplets, albeit to an extent that remains to be shown. We investigated this by conducting defogging experiments on four different samples. As before, the control surface was plain untreated glass (untreated control), which we initially compared to plain glass coated with poly(perfluorodecyl acrylate) (pPFDA; hydrophobic control) to render it hydrophobic (see Methods, section "Surface preparation" for further details). The third and fourth samples consisted of an untreated metasurface and a metasurface treated with pPFDA (hydrophobic metasurface) to render it hydrophobic (see Supporting Information, section "Advancing and receding water contact angles of the tested surfaces" and Figure S1 for details on wetting behavior). We initially exposed each surface to supersaturated conditions by cooling the surface. This caused a layer of condensate to form. The samples were then transferred to the experimental setup depicted in Figure 3a. To better mimic real-world conditions, we changed the power density to 1 sun (1000 W m$^{-2}$). We then measured the evaporation time for seven different samples of each sample type (see Methods, section "Setup and experimental protocol" for further details). We found that both untreated and hydrophobic metasurfaces outperformed the untreated and hydrophobic control samples by reducing the time necessary for complete surface defogging.

Figure 4a–b show image sequences of the untreated control and the untreated metasurface, respectively. Qualitatively, one observes that the condensate surface coverage remains similar for the first 70 s, followed by a faster reduction of condensate surface coverage on the untreated metasurface, and hence a reduced defogging time. Figure 4c shows a box plot of the defogging time ($t_{df}$) for seven individual runs on the two sample types depicted in Figure 4a–b. Figure 4c additionally shows $t_{df}$ for the hydrophobic control and metasurface. As expected, the quantitative analysis indicates that there is a clear improvement in $t_{df}$ for the metasurfaces when compared to the plain glass, illustrated by a two sample t-test p value < 0.001, indicating that the difference between the two population means is statistically significant. Here we note that the untreated metasurfaces have a slightly lower receding contact angle, by approximately 10°, than the untreated control surfaces (see Supporting Information, section "Observed effect of surface wettability on fogging resistance for further details" and Figure S1). However, the significant reduction of $t_{df}$ should not be attributed to this small difference in wettability and can only be due to the metasurface heating and corresponding evaporation enhancement by increasing the saturation vapor pressure above the droplets.

Additionally, our results in Figure 4c show that the hydrophobic surfaces have better performance in comparison to their respective untreated samples, both of which have a two sample t-test p value < 0.05. On the one hand this seems surprising because it is well known that droplets on a surface with a larger receding angle have a slower evaporation rate than neutrally wetting or hydrophilic droplets,[39–42] at least in the µL volume range. On such scales it is probable that, independent of the macroscopic receding angle, the droplets evaporate in a constant contact radius (see Figure 3b, the tail end of contact angle progression; see also Ref.[40]). On the other hand, we observed that the volume of liquid on the surface in the form of condensate droplets is approximately 10% less for hydrophobic glass than for the control glass (see Supporting Information, section "Observed effect of surface wettability on fogging resistance"). This means that there is simply less water on the surface to vaporize and offers the most rational explanation as to why we observed a decrease in $t_{df}$ for the hydrophobic samples.

As mentioned, the ability to inhibit fogging is an important quality of our metasurfaces and could be used to an advantage in many applications, particularly for specialized outdoor eyewear, which can suffer from visibility loss from fogging even during sunny weather. We tested the efficacy of our metasurfaces by comparing them to control samples, both untreated and superhydrophobic, exposing each sample to the same supersaturated conditions. We created a supersaturated environment by modifying our experimental setup, illustrated in Figure 5a, to include a chamber housing the sample and including a water bath and a small fan. Additionally, the camera was tilted at a shallow angle with respect to the sample plane in order to avoid direct light expose to the sensor (see Methods, section "Setup and experimental protocol" for further details). The fogged surfaces appear brighter due to light scattering by the droplets, to which the slightly tilted camera with respect to the incident light is sensitive. Figure 5b–d show image sequences of the fogging behavior of the untreated control, superhydrophilic state-of-the-art, and untreated metasurface samples, respectively, under supersaturated conditions. The first, and perhaps most important observation is that the metasurface remains, for the most part, condensate free. Although one observes slight condensate bursts on parts of the surface, it quickly evaporates again and at steady state the metasurface remains condensate free (see Supporting Information, Video S1). In contrast to the metasurface, the untreated control and superhydrophilic state-of-the-art surfaces in Figure 5b–c show immediate condensate formation. On the one hand, the condensate on the untreated control takes the form of many droplets that scatter the light and almost completely eradicate the visibility of the background. On the other hand, the condensate on the superhydrophilic state-of-the-art surface results in a film of water that distorts the background. Both of these surfaces significantly reduce the visibility retention of the background in comparison to the metasurface. Remarkably, the metasurface remains condensate free. Even though the metasurface is able to more efficiently turn the broadband illumination into heat and significantly raises the temperature of the surface in comparison to the control samples, the approximately 3 °C temperature increase due to the heat response of the metasurface combined with the convective heating inside the chamber produces a $p/p_\text{L}$ of approximately unity (1.01 ± 0.01) for the aforementioned experimental conditions. The control sample also profits from the convective heating inside the chamber, however the heat response from the light irradiance is considerably less resulting in a $p/p_L$ of 1.22 ± 0.01 and the necessary conditions for fogging to occur.

## Conclusions

Surface fogging is a common hindrance, especially when considering the clarity and distortion of light travelling through transparent or back from reflective surfaces. Due to the importance of the phenomenon there are a variety of existing approaches to mitigate this loss in optical efficiency, however we demonstrated that the most widely used solution, superhydrophilic surfaces, does not retain all optical efficiency aspects, resulting in significant image distortion, while untreated and hydrophobic surfaces result in significant loss of clarity. Both of these methods change the inherent surface energy of the substrate, which continues to prove to be a difficult characteristic to maintain for long periods. We demonstrated a novel approach by employing plasmonic metasurfaces harnessing the broadband spectrum of the sun, to efficiently heat up a surface, thereby significantly improving defogging and anti-fogging properties, without a marked loss of transparency. For applications that require reduction of light transmission, the broadband absorption of our metasurfaces offers a favorable property compared to commercially available products used for reducing light transmission, such as tinted laminates. This approach would result in considerable performance gains for applications such as windows, windshields, electronic displays, cameras, mirrors, and eyewear. The variety of substrates, including polymers and flexible surfaces, which can be coated with our metasurfaces as well as the potential manufacturing scalability of the metasurface coating process make this technology a viable solution for anti-fogging and defogging in commercial applications.

## Methods

**Surface preparation.** We used double side polished, 4-in, 500-μm thick wafers (UniversityWafer, Inc.) made of fused silica. We cut the wafers into 5 mm x 5 mm square specimen, using a wafer dicer (ADT

ProVectus LA 7100) with a diamond head. Prior to cutting, a 3.3 µm polymer top-layer was applied to protect the glass surface. After the dicing process, the protective layer was removed by sonication in acetone for 2 min, followed by an equal-time sonication in isopropyl alcohol, and water. We also treated each specimen with oxygen plasma (Oxford Instruments, Plasmalab 80 Plus) for 7 min. Shortly after, we prepared 7 different sample types:

a. *Untreated control*. We stored freshly-diced and cleaned glass wafer pieces in a dust-free environment, under ambient temperature (≈23 °C) and relative humidity (RH = 40 – 60 %), for several days (>5 d), enough time for the adsorbed hydrocarbon layer to approach equilibrium with the surfaces (therefore providing stable advancing and receding contact angles).

b. *Hydrophilic control*. We immediately (<1 hr) used the plasma-treated glass specimen in the clarity and distortion evaluation, as well as in the defogging experiments.

c. *Untreated metasurface*. We used sputter deposition (Von Ardenne CS 320 C) to fabricate the light-absorbing metasurface on top of the fused silica substrates, in line with previous research.[43] Two component materials were used, gold nanoparticles (Au) and titanium dioxide ($TiO_2$), with the following deposition parameters: 50 W DC field, 6 µbar pressure, 3 s duration, and 600 W RF field, 6 µbar pressure, 43 s duration, respectively. We fabricated a structure of 8 alternating Au and $TiO_2$ layers in total, with the first layer being $TiO_2$. The metasurface was finalized with an extra $TiO_2$ top-layer (deposition time of 72 s). The thickness of the metasurface has been estimated ≈60 nm, with a mean absorption of 37% and transmission of 36% (see also Figure 2) in the wavelength range of 400 – 750 nm.

d. *Silane-treated control*. For the single droplet evaporation experiments, we used a homebuilt vacuum chamber for physical vapor deposition of trichlorovinylsilane (TCVS; 97%, Sigma-Aldrich) on our glass substrates. For this purpose, we first transferred 1 mL of TCVS into a glass container, under nitrogen atmosphere. With the container connected to the chamber and its valve closed, we then evacuated the chamber until a stable pressure of 100 µbar was reached, at which moment the valve was opened, while the chamber was also isolated from the vacuum pump. Deposition time was 20 min, at a temperature of

≈23 °C. In the end, we removed the excess of TCVS with a prolonged evacuation and purging cycle.

e. *Silane-treated metasurface*. We followed the same process as in (d) to deposit TCVS on our glass specimen already coated with metasurfaces.

f. *Hydrophobic control*. We utilized initiated chemical vapor deposition method (iCVD; iLab™ Coating System, GVD Corporation) to deposit ultrathin conformal films of poly(perfluorodecyl acrylate) (pPFDA) on clean glass substrates, in order to render them hydrophobic. The advantage of pPFDA over other fluoropolymers lies on its low surface energy (9.3 mN/m), which is half of the surface energy of the widely-used poly(tetrafluoroethylene) (18 mN/m).[44] Silane-treatment of the substrates according to (d) precedes iCVD. The covalently bound surface vinyl groups in TCVS react with the vinyl groups in perfluorodecyl acrylate monomers and bind them chemically to the substrate.[45] For the iCVD deposition we used 1H,1H,2H,2H-perfluorodecyl acrylate (PFDA; 97%, with *tert*-butylcatechol as inhibitor, Sigma-Aldrich) and *tert*-butyl peroxide (TBPO; 98%, Sigma-Aldrich) as monomer and initiator, respectively.[46] The main deposition parameters are reported here: TBPO flow rate was 2.6 sccm, PFDA flow rate was 1.0 sccm, substrate temperature was ≈20 °C, filament temperature was ≈300 °C, PFDA container temperature was 80 °C, process pressure was ≈100 mbar, and deposition time was 15 min. We stabilized all temperatures for 5 min prior to each deposition; we also purged the vacuum chamber with nitrogen for 10 min after deposition.

g. *Hydrophobic metasurface*. We followed the same procedure as in (f) to deposit pPFDA on our glass, metasurface-coated specimen.

**Surface characterization.** The cross section of the metasurface was imaged by a Hitachi SU8230 scanning electron microscope, while the top-view image was acquired with a Brooker Dimension Fastscan atomic force microscope. We measured the reflection and transmission spectrum of the metasurface, tinted laminate (LLumar Esprit Series ATC 35 CH SR HPR), and control samples for the 400–1650 nm wavelength range using a home-built spectroscopic system.[47] We measured the advancing and receding contact angles of each sample type using a DataPhysics OCA 35 goniometer and their SCA software, by the inflation/deflation method (droplet volume of 2–3 µL). The observations of the condensing droplets under the microscope used for the Supporting Information were performed using an upright microscope (Olympus BX60), with a 5x objective (Olympus MPlan 5x/0.10 BD JAPAN) and a 20x objective (LMPLanFI 20x/0.40 BD JAPAN) in bright-field configuration.

**Setup and experimental protocol.** We carried out the clarity and distortion measurement experiments using an upright microscope (Olympus BX60) and a 5x objective (Olympus MPlan 5x/0.10 BD JAPAN) in dark-field configuration. We placed the matrix of dots, which served to quantify the clarity and distortion, on top of a Peltier element. We then placed the sample in question on top of the dotted matrix and blew a steady stream of nitrogen onto the sample while cooling it down to 6 °C with the peltier element in order to create a locally dry environment, to prevent premature condensate formation on the sample. After the temperature had stabilized, we stopped the nitrogen stream, exposing the sample to 40% relative humidity at 24 °C ($p/p_L = 1.28$), at time $t = 0$ s. We then recorded the fogging process on the sample exposed to the abovementioned supersaturated conditions for a duration of 10 min ($t = 600$ s).

We conducted the single droplet evaporation experiments using the setup depicted in Figure 3a. Using two lenses, we collimated and focused a broadband (sun-mimicking) halogen light source (Schölly Fiberoptic GmbH, Flexilux 600 longlife 1.25 A) onto each test sample at a power density of 4000 W m$^{-2}$ (corresponding to 4 suns). In order to get the most representative results for the aforementioned applications, it was important that the surface of each sample was fully irradiated by the focused light source and that as little contact with the holder as possible was made in order to limit heat diffusion from the metasurface to the surrounding media. Our focused light had a diameter of $\approx 7$ mm and we used 5 mm x 5 mm samples. We irradiated the sample with light until its temperature reached steady state (see Supporting Information, section "Infrared thermography and temperature response measurements"). We recorded the temperature of the sample using an infrared (IR) camera (FLIR SC7500, spectral range of 1.5–5.1 µm) before the droplet was placed on the sample ($t < 0$ s) as well as after placing the droplet and throughout its complete evaporation period. After placing an approximately 1 µL droplet on the sample, we additionally began recording a side view using the CMOS camera. In order to allow each droplet to reach a constant receding contact angle and therefore give comparable data between sample types, $t = 0$ s corresponds to the moment when each droplet reached a volume of 0.6 µL. Starting at $t = 0$ s, we computed the contact angle and volume of the droplet until it had completely evaporated. For both sample types (untreated control and untreated metasurface), we carried out three experiments on different samples. For the IR images in Figure 3d–e we note that images of the surface and the droplet were separately calibrated (due to differing emitted signal for the same temperature) and overlaid in order to enable usage of a single temperature scale bar. We carried out all the experiments in an environment with RH = 64%, at 21.8 °C. We also acquired the transient thermal response of an untreated control, an untreated metasurface, and a commercial absorber (Thorlabs, NE504B; unmounted absorptive neutral density filter; optical density of 0.4) using the infrared camera, with a 50 mm F/2 lens, at a framerate of 10 fps, and for two different power densities (1 sun and 4 suns). We kept the optical axis of the camera at a 5° angle with respect to the out of plane vector of each surface under test in order to alleviate the Narcissus effect.

For the multidroplet defogging experiments shown in Figure 4 we used the same setup as depicted in Figure 3a, mounting a Canon 5D Mark III DSLR camera at an angle of approximately 45°. We placed a fishbone pattern on the holder below the sample in order to improve contrast and illustrate the sample transparency. We fogged each sample by cooling it down to 0 °C for 5 min in an environment with a relative humidity of 60% at 23.0 °C ($p/p_L = 2.75$). We proceeded to take the sample off the peltier element and transferred it to the holder, which was then irradiated with light at a power density of 1000 W m$^{-2}$ (approximately 1 sun) in an environment with a relative humidity of 60% at 23.0 °C. We measured the time and visually recorded the samples using the camera until the condensate had completely evaporated. We conducted 7 experiments on different samples for each sample type (control untreated, control hydrophobic, metasurface untreated, metasurface hydrophobic). We tested each sample population for a standard normal distribution using the Kolmogorov-Smirnov test and validated the improvement in defogging time using the two-sample t-test.

We carried out the anti-fogging experiments using the setup depicted in Figure 5a. We placed a small printed illustration (edelweiss flower) below the sample holder, to give an enhanced impression of visibility retention. The illustration was sprayed with a passive anti-fogging agent (Aqualung Anti Fog Spray) to reduce condensation on the flower (not completely effective as condensate is slightly visible for $t = 30 - 60$ s in Figure 5d, behind the otherwise fog-free metasurface). The experiments were conducted by setting the temperature of the water bath ($T_w$) and measuring the environmental air temperature inside the chamber ($T_e$). We also measured the sample temperature ($T_s$) to determine $p/p_L$. We used the fan in order to get a homogeneous distribution of $p/p_L$ throughout the chamber. When $T_w$ reached 65 °C, we turned the fan on and the experiments began, corresponding to $t = 0$ s. We recorded the process for 60 seconds. In order to determine the supersaturation of the vapor in the chamber with respect to the sample we ran three individual experiments both for the metasurface and control samples. For the metasurface we obtained an average environmental temperature across the three experiments of 35.9 ± 0.3 °C and an average surface temperature of 35.8 ± 0.4 °C. These three experiments resulted in an average supersaturation above the metasurface of 1.01 ± 0.01. For the glass surface we obtained an average environmental temperature across the three experiments of 37.3 °C ± 0.5 °C (the increase in environmental temperature was due to the slow warming of the chamber after repeated experiments) and an average surface temperature of 33.7 °C ± 0.6 °C. These three experiments resulted in an average supersaturation above the glass surface of 1.22 ± 0.01.

## Supporting Information

The Supporting Information is available online. The following sections are included: "Advancing and receding water contact angles of the tested surfaces", "Theoretical effect of surface wettability on fogging resistance", "Observed effect of surface wettability on fogging resistance", and

"Infrared thermography and temperature response measurements" (PDF).

Video S1: "Anti-fogging behavior of metasurfaces".

## Acknowledgments

We thank J. Vidic and P. Feusi for assistance in experimental setup considerations and construction; U. Drechsler and D. Caimi for assistance in surface fabrication; and Dr. F. Dähler and Prof. A. Steinfeld for access and assistance with spectroscopy measurements. We acknowledge support by the European Research Council under Advanced Grant 669908 (INTICE).

## Author Contribution Statement

T.M.S., H.E., and D.P. designed the research and provided scientific advice on all of its aspects; C.W., E.M., and T.K. performed experiments and analyzed data; E.M. fabricated and characterized surfaces; C.W., E.M., T.M.S., and D.P. wrote the paper.

## Additional Information

**Competing financial interests:** The authors declare no competing financial interests.

# Figures and Captions

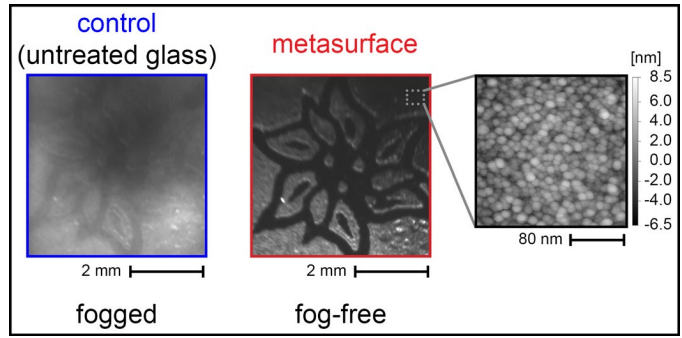

**Table of Contents Graphic. Transparent Metasurfaces Counteract Fogging.** Left and middle: Superior anti-fogging performance of our metasurfaces with respect to an untreated control glass substrate, after an exposure time of 30 s to supersaturated conditions, under sunlike illumination. Right: Atomic force micrograph of the gold nanoparticles, constituents of the metasurfaces.

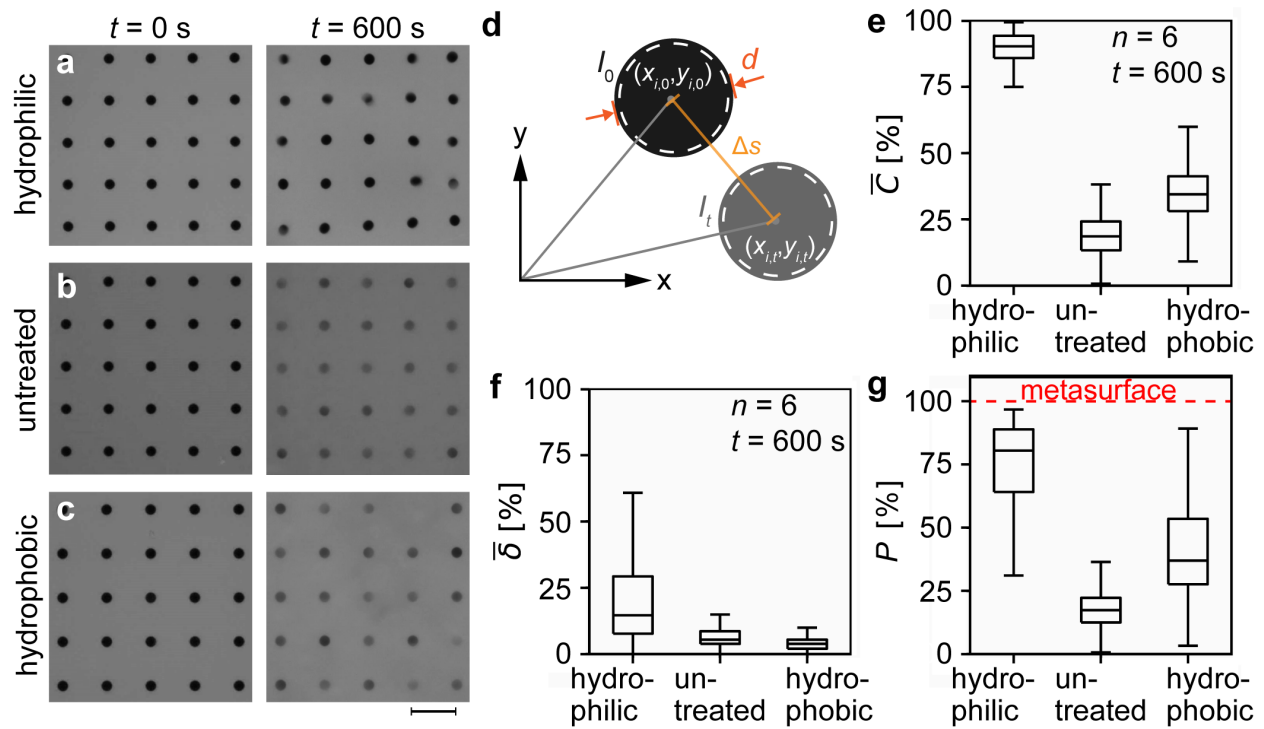

**Figure 1. Effect of fogging on the clarity and distortion for differing wetting behavior of transparent surfaces.** We characterized both the clarity and distortion of transparent glass surfaces by using a dot array placed under the surfaces. (a–c) Darkfield micrographs of the surfaces and the dot array were taken before ($t = 0$ s) and 10 min after ($t = 600$ s) the samples were exposed to supersaturated vapor conditions. The before and after for the (a) superhydrophilic, (b) untreated (neutrally wetting), and (c) hydrophobic glass are shown. We quantified the clarity and distortion by analyzing the dot array for each sample type as illustrated in (d) using Eqs. 1 and 2. The quantified effect of fogging on (e) clarity ($\bar{C}$) and (f) distortion ($\bar{\delta}$) on each of the three surfaces is illustrated as box plots, where each type of surface was tested using 6 different samples ($n=6$). We also estimated the visibility retention through the three surfaces by introducing a performance factor, $P$, and compared them to a sunlight absorbing metasurface ($P=1$), as shown in boxplot (g). Scale bar (a–c): 100 µm.

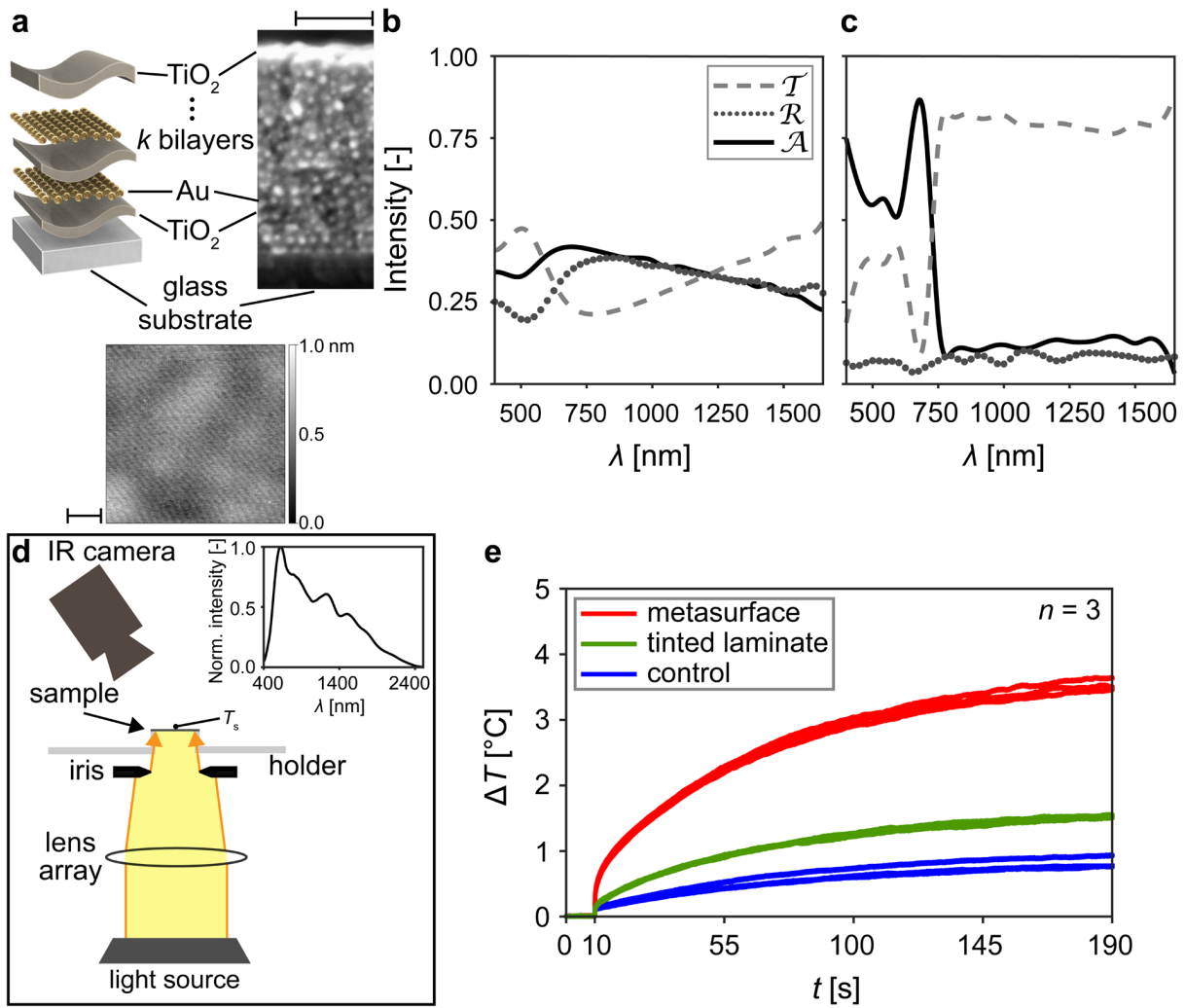

**Figure 2. Topographical, optical and light-induced heating characterization of the metasurface and comparison to commercially available tinted laminate with similar transmittance in the visible spectrum.** (a) Top left: Schematic representation of the multilayer structure of Au nanoparticles and TiO₂, which make up the metasurface; top-right: scanning electron micrograph of the metasurface cross-section, acquired with secondary electrons, where the bright circular spots are the Au nanoparticles; bottom row: atomic force micrograph of the metasurface. (b–c) Reflectance ($\mathcal{R}$), transmittance ($\mathcal{T}$), and absorbance ($\mathcal{A}$) spectra of the (b) metasurface and (c) commercially available tinted laminate for visible and near infrared wavelengths (400–1650 nm). (d) Experimental setup showing how the temperature increase of the samples upon illumination was measured. (e) Sample temperature increase, $\Delta T$, over time, $t$, for the metasurface, tinted laminate, and control glass samples. Sample thickness was $\approx 4$ mm. Scale bars (a): 50 nm.

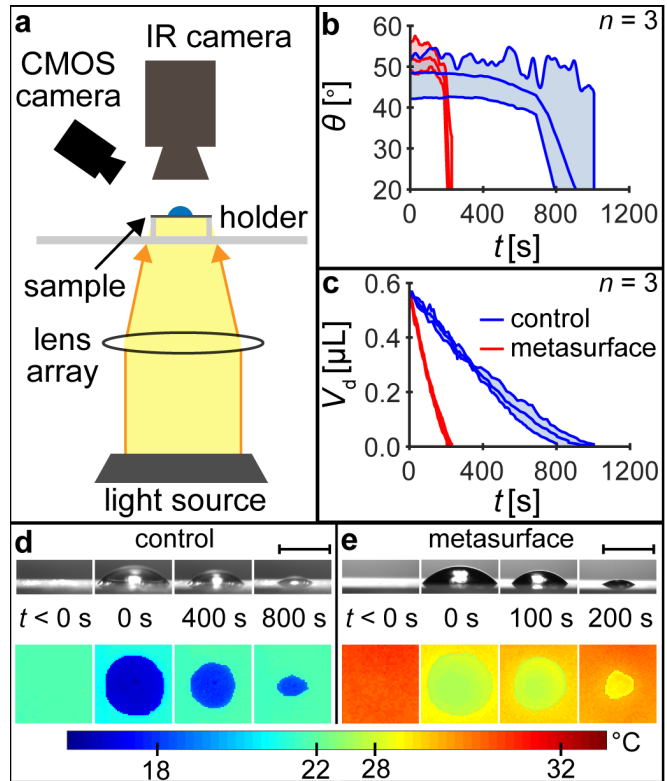

**Figure 3. Sunlight absorbing metasurfaces drastically reduce evaporation time of a single droplet.**
**(a)** Experimental setup consisting of a sun-mimicking light source and a lens system to focus the light on the sample with a power density of 4000 W m$^{-2}$. We used an angled side-view CMOS camera to observe the contact angle progression of the droplets and a top-view IR camera to record the sample and droplet temperature profiles. **(b)** Plot of contact angle, $\theta$, vs time, $t$, for control samples (red curves) and metasurfaces (blue curves). The evaporation rate of the metasurfaces was approximately four times higher than the control, as seen in plot **(c)**, droplet volume, $V_d$, vs $t$. **(d–e)** Side view and infrared top view of a single droplet evaporating on a **(d)** control surface and **(e)** metasurface. Scale bars **(d–e)**: 1 mm.

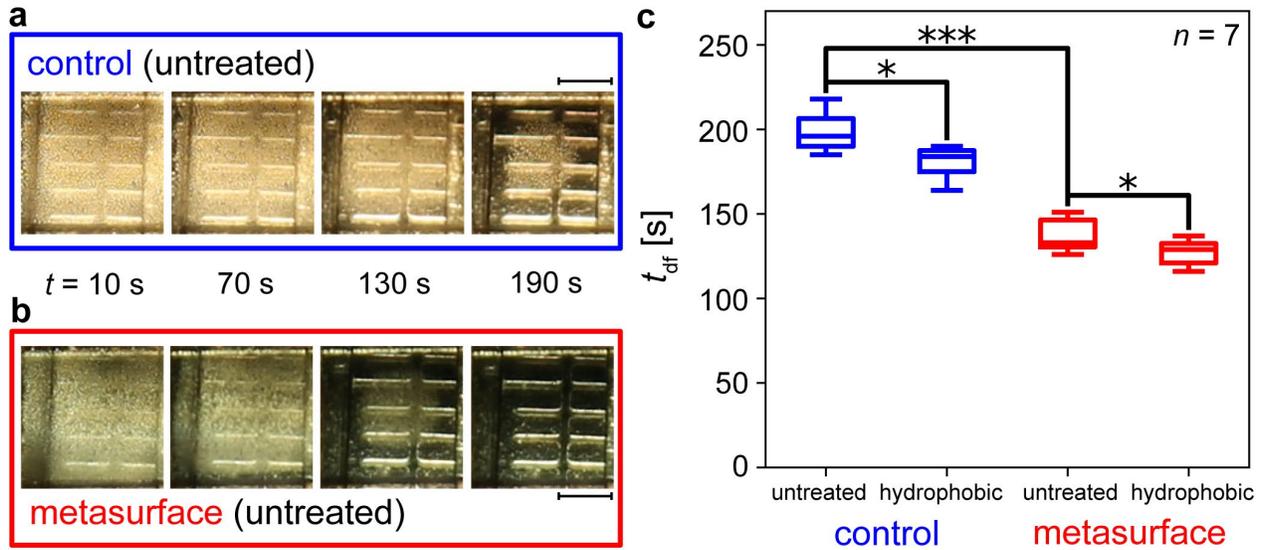

**Figure 4. Using metasurfaces to enhance visibility restoration by reduced defogging time.** Using metasurfaces that efficiently turn the absorbed light into heat, we were able to reduce the defogging time compared to a control surface. The transparency vs $t$ is shown for an **(a)** untreated glass surface (untreated control) and an **(b)** untreated glass metasurface (untreated metasurface). Statistically, the metasurfaces reduced the defogging time ($t_{df}$) by approximately 25% as shown in **(c)**. We tested both the control and metasurface samples against their hydrophobic counterparts to see if hydrophobicity also has an effect on $t_{df}$. For both pairs of samples we obtained two sample t-test p values < 0.05 to prove this was the case. Significance bars in **(c)** are represented by * for $p < 0.05$ and *** for $p < 0.001$. Scale bars **(a–b)**: 2 mm.

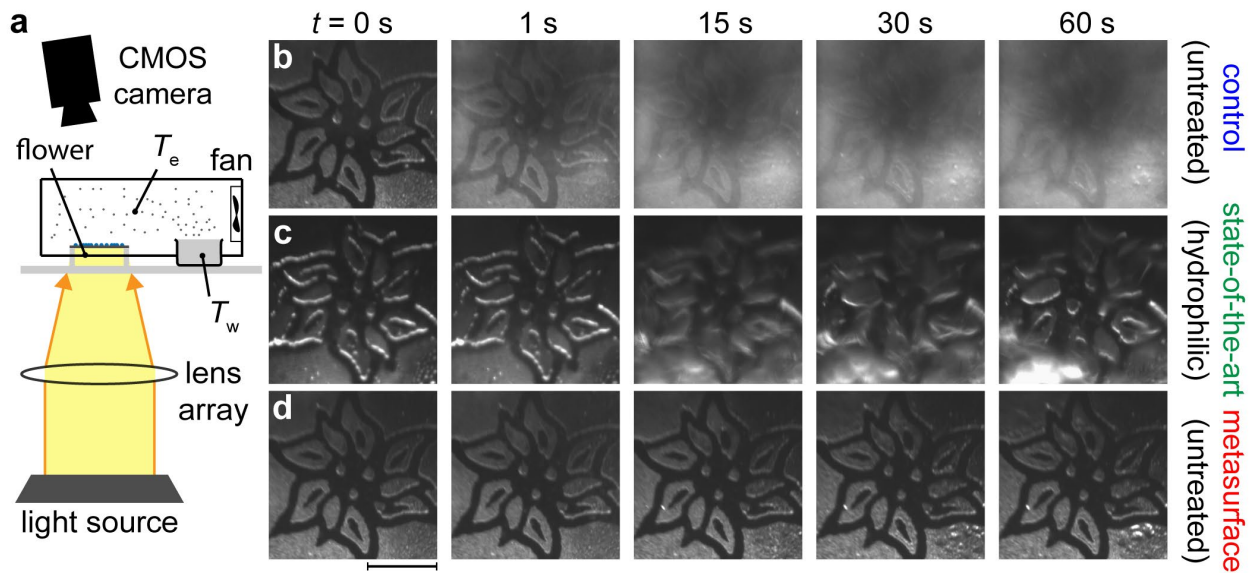

**Figure 5. Anti-fogging performance enhancement by efficient sunlight absorbing metasurfaces.** We tested the anti-fogging performance of our metasurfaces and compared it to the performance of untreated control and superhydrophilic state-of-the-art samples, using the setup illustrated in **(a)**. We enhanced the setup with the inclusion of a simple environmental chamber in which we placed a bath of warm water at a temperature $T_w$ while also measuring the environmental temperature ($T_e$) and the sample temperature ($T_s$), in order to achieve a supersaturated environment. We used a fan to ensure that the water vapor inside the chamber was well mixed. The results for 60 s of supersaturation are shown in the image sequences for the **(b)** untreated control, **(c)** superhydrophilic state-of-the-art, and **(d)** untreated metasurface samples. In order to better demonstrate the distortion effect of the superhydrophilic sample, we slightly increased the distance between the surface and the flower illustration. Scale bar **(b–d)**: 2 mm.

# Supporting Information

## Transparent Metasurfaces Counteracting Fogging by Harnessing Sunlight

*Christopher Walker†, Efstratios Mitridis†, Thomas Kreiner, Hadi Eghlidi, Thomas M. Schutzius\*, Dimos Poulikakos\**

Laboratory of Thermodynamics in Emerging Technologies, Department of Mechanical and Process Engineering, ETHZ Zurich, Sonneggstrasse 3, CH-8092 Zurich, Switzerland

**Keywords: Antifogging, defogging, condensation, renewable energy, metasurface**

† denotes equal contribution

\* To whom correspondence should be addressed

Prof. Dimos Poulikakos
ETH Zurich
Laboratory of Thermodynamics in Emerging Technologies
Sonneggstrasse 3, ML J 36
CH-8092 Zurich
Switzerland
Phone: +41 44 632 27 38
Fax: +41 44 632 11 76
dpoulikakos@ethz.ch

Dr. Thomas M. Schutzius
ETH Zurich
Laboratory of Thermodynamics in Emerging Technologies
Sonneggstrasse 3, ML J 27.2
CH-8092 Zurich
Switzerland
Phone: +41 44 632 46 04
thomshu@ethz.ch

## Advancing and receding water contact angles of the tested surfaces

Both the advancing ($\theta_a^*$) and receding ($\theta_r^*$) contact angles for each of the tested surfaces used in this study are an integral part that we use to explain the phenomena that we observed. We measured and summarized both $\theta_a^*$ and $\theta_r^*$ in Figure S1 for all of the surfaces used in this study, using 3 different samples of each kind and 2–3 measurements for each sample, for a total of 7 measurements ($n=7$). Figure S1a,b show the spread across the measurements for each sample type using box and whisker plots. For our study it is important that $\theta_a^*$ and $\theta_r^*$ are comparable between the same sample types of control and metasurface (i.e. the silane-treated control and metasurface must have similar contact angles). This is important in order to rule out any sort of effect on antifogging or defogging that is not due to the sunlight absorbing properties of the metasurfaces.

Inspecting Figure S1, one observes that, in spite of the fact that surface chemistry is expected to be identical between metasurfaces and control samples of the same type, $\theta_a^*$ is slightly larger and $\theta_r^*$ is slightly smaller for the metasurfaces. This increase in the contact angle hysteresis is a typical signature for increased surface roughness, assuming that the droplets sit in the Wenzel state.

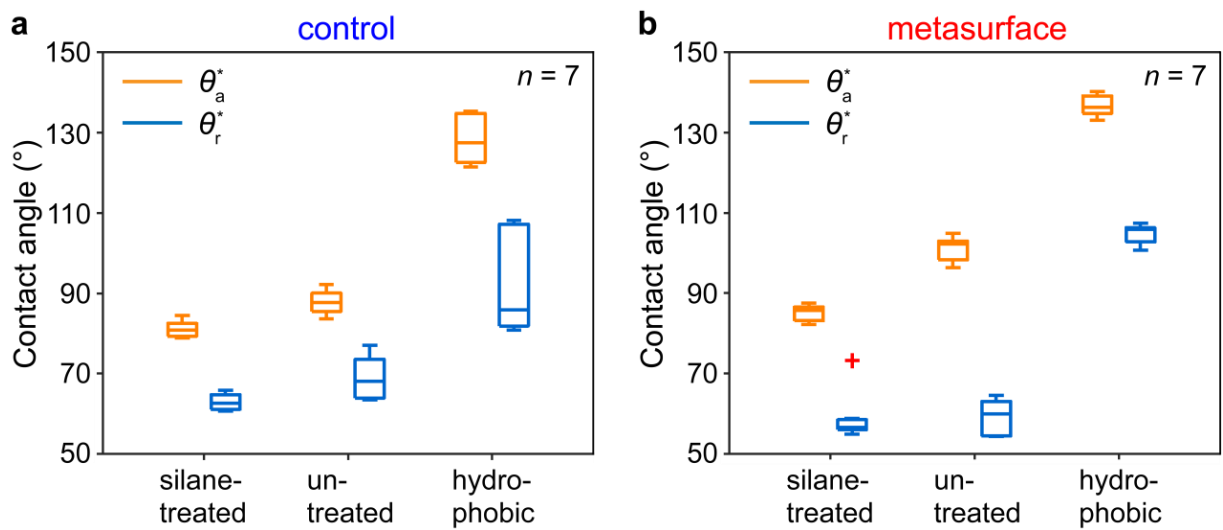

**Figure S1. Advancing and receding contact angles of the control samples and metasurfaces used in this study. (a–b)** Advancing, $\theta_a^*$, and receding, $\theta_r^*$, contact angles of the following samples types: treated with trichlorovinylsilane (silane-treated), untreated and treated with poly(perfluorodecyl acrylate) (hydrophobic), for **(a)** control and **(b)** metasurfaces.

## Theoretical effect of surface wettability on fogging resistance

Using classical nucleation theory one can derive the supersaturation ($p/p_L$) at which heterogeneous nucleation of liquid water from water vapor should theoretically occur on a surface. This is important for antifogging applications as it provides design rules in order to engineer superior fog-resistant materials. We begin the derivation by defining the critical free energy necessary to overcome in order to form a nascent water embryo in the absence of any surface ($\Delta G^*$, homogeneous nucleation) as:

$$\Delta G^* = 16\pi\gamma^3 \Big/ \left(3\left[n_L kT \ln(p/p_L)\right]^2\right) \tag{S1},$$

where $\gamma$ is the surface tension of water in air, $n_L$ is the number of water molecules per unit volume, $k$ is Boltzmann's constant, at temperature $T$ and supersaturation $p/p_L$, where $p$ is the water vapor pressure of the environment and $p_L$ is the saturated vapor pressure over a plane surface of water. Because of the inverse squared relationship of $p/p_L$ to the natural logarithm, a larger $p/p_L$ results in a significantly lower $\Delta G^*$.

The critical free energy for nucleation on a surface is a modification of this equation by a factor ($f$), which is a function of the water contact angle on the surface ($\theta$), assuming the surface is flat, given by:[1]

$$\Delta G^* = 16\pi\gamma^3 \Big/ \left(3\left[n_L kT \ln(p/p_L)\right]^2\right) f(\theta) \tag{S2}.$$

The function $f(\theta)$ can be expressed as:[1]

$$f(\theta) = 1/4\left(2 - \cos\theta + \cos^3\theta\right) \tag{S3},$$

where $0 < f(\theta) < 1$; a larger $\theta$ results in larger $f(\theta)$. This also means that a larger $\theta$ results in a higher $\Delta G^*$. The final equation that gives us the nucleation rate ($J$), which allows us to predict the probability of a nucleation event is given by:

$$J = J_0 \exp\left(-\Delta G^*/kT\right) \tag{S4},$$

where $J_0$ is a rate constant determined by the rate at which vapor molecules strike a unit area of the embryo. Here, it is important to observe that $J$ has an inversely exponential relationship to $\Delta G^*$ and therefore will grow exponentially larger as $\Delta G^*$ becomes smaller.

Using equations (S2), (S3) and (S4), Figure S2 illustrates the dependence of $J$ on $p/p_L$ and $\theta$. We observe the exponential dependence of $J$ on $p/p_L$ for both $\theta$. More importantly, however, we observe that the difference in $\theta$ changes $J$ by many orders of magnitude. For example, at $p/p_L \approx 2.25$, a surface with $\theta = 80°$ would result in $J = 1$ mm$^{-2}$ s$^{-1}$, while a surface with $\theta = 125°$ at the same $p/p_L$ would result in $J \approx 10^{-35}$, or for all intents and purposes no observation of nucleation. Furthermore, $p/p_L$ would have to be $\approx 3.5$ before a surface with $\theta = 125°$ would result in a substantial nucleation rate $J = 1$ mm$^{-2}$ s$^{-1}$. Differences in surface wettability have been shown to create favorable condensation nucleation on more hydrophilic areas,[2] however a quantified difference of $p/p_L$ between the two surfaces was not measured. Nonetheless, this analysis indicates that one could theoretically engineer more fog-resistant surfaces by simply increasing the surface hydrophobicity in order to inhibit the formation of condensate on the surface. Of course, after the condensate has been formed, the influence of the contact angle of the microdroplets on light scattering is another aspect that must be considered, especially for large droplets or high water surface coverage.

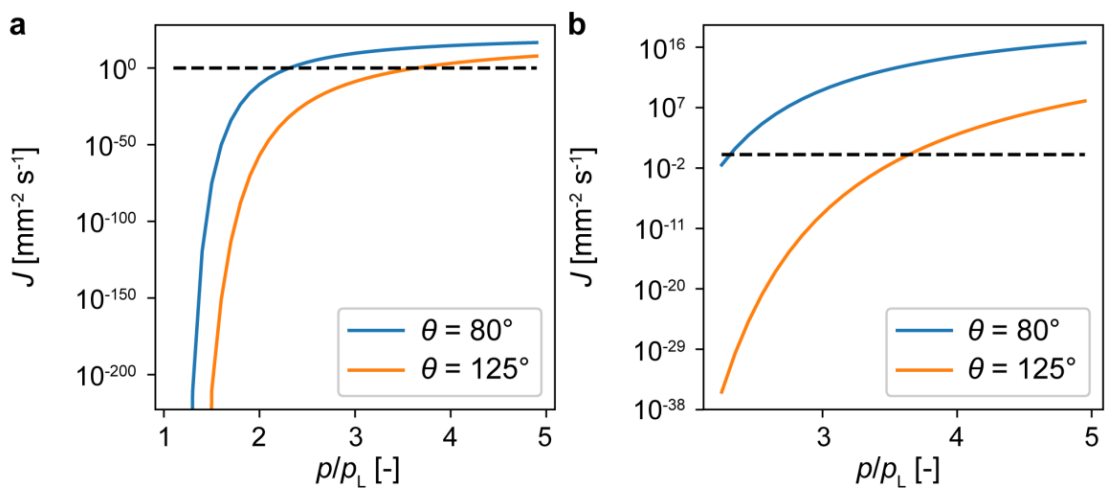

**Figure S2. Dependence of the heterogeneous nucleation rate ($J$) on both vapor supersaturation and water contact angle.** Due to the increased surface area of higher contact angle droplets the energy barrier for heterogeneous nucleation ($\Delta G^*$) increases with contact angle. The exponential dependence of $\Delta G^*$ on $J$ is illustrated in **(a)**. This relationship can be better observed in **(b)**: zoomed in plot showing the range of supersaturations in which $J$ becomes substantial, illustrating that $J$ for a substrate with $\theta = 125°$ is $>30$ orders of magnitude lower, as compared to the $p/p_L$ where $J=1$ mm$^{-2}$s$^{-1}$ for a substrate with $\theta = 80°$. The black dashed line corresponds to $J = 1$ mm$^{-2}$s$^{-1}$.

## Observed effect of surface wettability on fogging resistance

It has been shown that a biphilic surface exposed to supersaturated conditions nucleates preferentially on the hydrophilic areas. Furthermore, classical nucleation theory predicts that hydrophobic surfaces should be able to also withstand much higher supersaturation without nucleation events when compared to hydrophilic surfaces. Remarkably, we found that the hydrophobic surfaces were not able to withstand higher supersaturation than neutrally wetting surfaces, rather they simply reduced the number of nucleation sites. Figure S3a,b show a control glass surface and a hydrophobic surface, respectively, that have been exposed to slightly supersaturated ($p/p_L \approx 1.02$) conditions for 60 s. Based upon the aforementioned plot in Figure S2a, we would expect an extremely low $J$ for both surfaces, however experimentally we still observe nucleation on both surfaces. This may be best explained by either contamination or surface adsorption of hydrocarbons and other environmental contaminants, both of which may render a hydrophobic surface more hydrophilic. Furthermore, surface cavities also increase the likelihood of nucleation. Since we observe more nucleation sites on the control surface than on the hydrophobic surface, it begs the question why more foreign material acting as nucleation sites would be adsorbed to the control surface. We suggest that the high surface energy of highly hydrophilic materials causes them to attract more adsorbates.[3]

Although hydrophobic coatings may reduce the nucleation density, they do not solve the problem of retaining visibility of transparent surfaces under supersaturated conditions (refer to Figure 1c). Nonetheless, they reduce the total amount of condensate coverage on the surface as compared to a more hydrophilic surface. We measured the volume of water on the surface using the droplet sizes, assuming an advancing contact angle and using the spherical cap approximation, for our hydrophobic sample, and compared it to the corresponding water volume on an untreated sample. We calculated a reduction of approximately 10% in water volume for the hydrophobic surface, compared to the untreated one, after 60 s of exposure to supersaturated conditions.

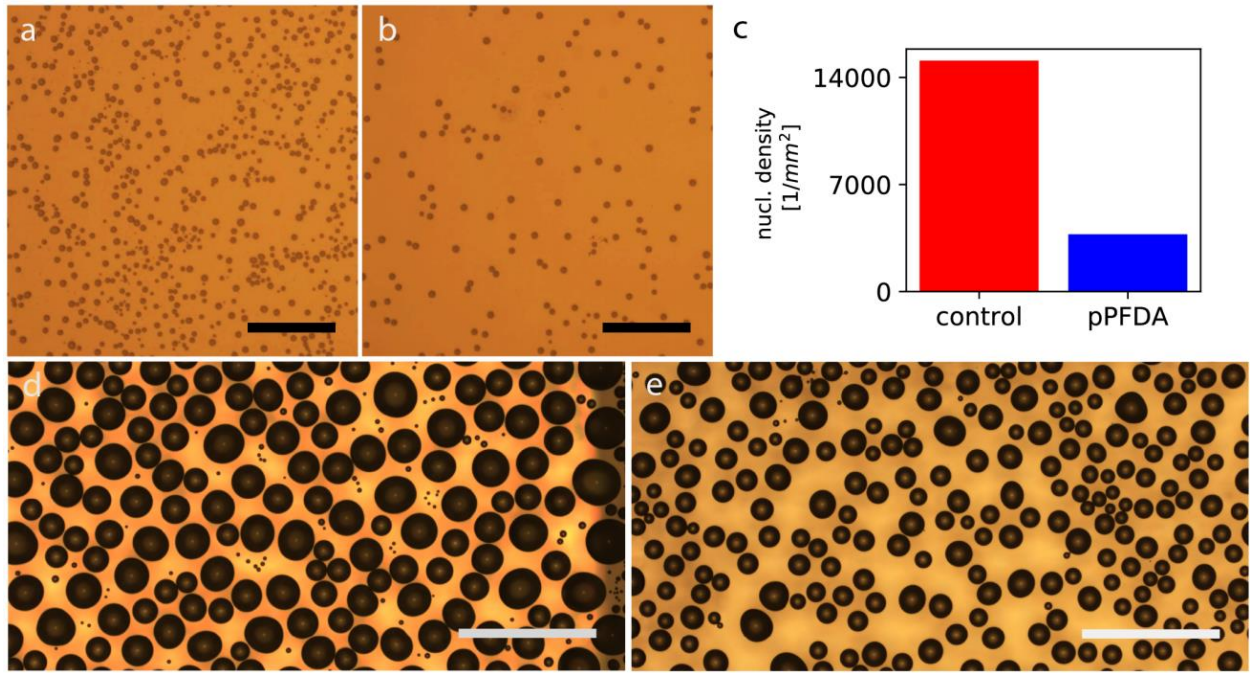

**Figure S3. Effect of the water contact angle on condensation nucleation density.** We compared the nucleation density of the untreated glass (control) and the pPFDA-coated glass. Micrographs recorded 60 s after the onset of nucleation on **(a)** untreated glass and **(b)** pPFDA-coated glass show that the nucleation density is higher on the untreated glass in comparison to the pPFDA-coated glass, although nucleation is observed to happen at the same supersaturation (just above supersaturation). The advancing and receding contact angles of the untreated glass are $\theta_a^* = 87° \pm 3.0°$ and $\theta_r^* = 68.7 \pm 5.5°$, respectively, while the advancing and receding contact angles of the pPFDA-coated glass are $\theta_a^* = 128.5° \pm 6.2°$ and $\theta_r^* = 93.3 \pm 13.3°$, respectively, indicating that the pPFDA-coated surface is considerably more hydrophobic. According to classical nucleation theory, the barrier for condensation nucleation should be considerably higher for the more hydrophobic surface. Using computer-aided image analysis, the **(c)** nucleation density on each substrate was quantified. The nucleation density has an influence on the condensate coverage after exposure to a supersaturation of 1.3 for 10 min as illustrated for the **(d)** control sample and **(e)** pPFDA-coated glass. The pPFDA-glass has approximately 10% less water on its surface, as compared to the control sample. Scale bars: (a–b), 50 µm; (e–f), 500 µm.

**Infrared thermography and temperature response measurements**

Infrared thermography is a non-intrusive surface temperature measurement technique, especially useful in cases where the thermal mass of the surface in test is very low. This way, we ensure that the actual temperature of our samples upon illumination is measured, while it also enables contactless observations during the single droplet evaporation experiments shown in Figure 3. A simplified expression of the temporally and spatially changing signal that the infrared camera receives can be written as:

$$S_t = \varepsilon S_s + S_{base} \quad (S5),$$

where $\varepsilon$ is the sample emissivity, $S_s$ is the signal that the sample would emit if it was a blackbody ($\varepsilon = 1$) and $S_{base}$ is the residue background signal.[4] Using equation (S5) for two discrete temperatures, ambient, $T_{amb}$ and an elevated one, $T'$, and subtracting in parts, assuming constant background signal, we have:

$$S_t^{T'} - S_t^{T_{amb}} = \varepsilon \left( S_s^{T'} - S_s^{T_{amb}} \right) \quad (S6).$$

We calibrated the infrared camera independently for the metasurface, control, and tinted laminate by using a hot plate and covering a temperature range of approximately 20 °C above the ambient. This way, $\varepsilon$ could be estimated ($\varepsilon \approx 0.9$ for the control and metasurface), and signal was converted into temperature by using the internal calibration curve of the camera, $S_s^{T'} = f(T)$. In the IR images shown in Figure 3d,e, a standard emissivity of $\varepsilon = 1$ was assumed specifically for the water droplets.

Figure S4a shows temperature increase *vs* time for a 500 μm untreated control and an untreated metasurface due to illumination, using a visible (halogen) light source and a power density of 1 sun for the illumination. We ran 3 experiments for each sample type. Figure S4b shows the temperature response curves at an increased power density of $\approx 4$ suns. Here we must mention that those temperatures were acquired by drawing a square box which covered >75% of the total sample area (5 mm x 5 mm) and taking the average temperature of the square for each time moment. The maximum temperature increase (180 s after the light was switched on) was 0.2 °C and 0.3 °C for the bare control, and 3.8 °C and 15.4 °C for the bare metasurface, at a power density of 1 sun and 4 suns, respectively.

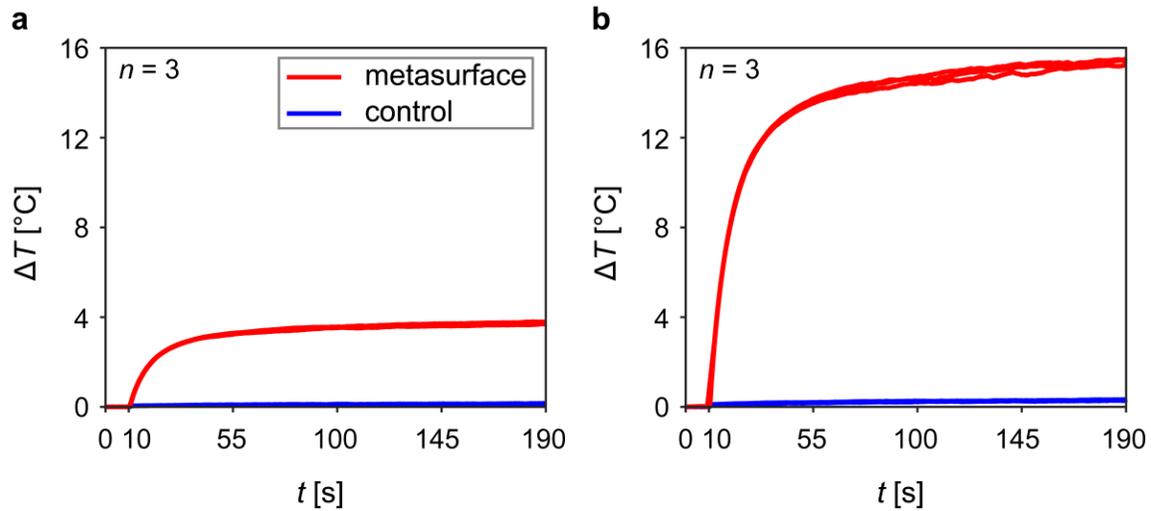

**Figure S4. Infrared-measured temperature response of water droplet-free untreated control (glass) samples and our metasurfaces, upon sun-like illumination.** Temperature increase with respect to ambient, $\Delta T$, vs $t$, for control and metasurface, at a power density of **(a)** 1 sun (1000 W m$^{-2}$) and **(b)** 4 suns. The sample size was 5 mm by 5 mm. The illumination was switched on at $t=10$ s, and was kept on for 180 s. The ambient temperature was $23.5\pm0.5$ °C.

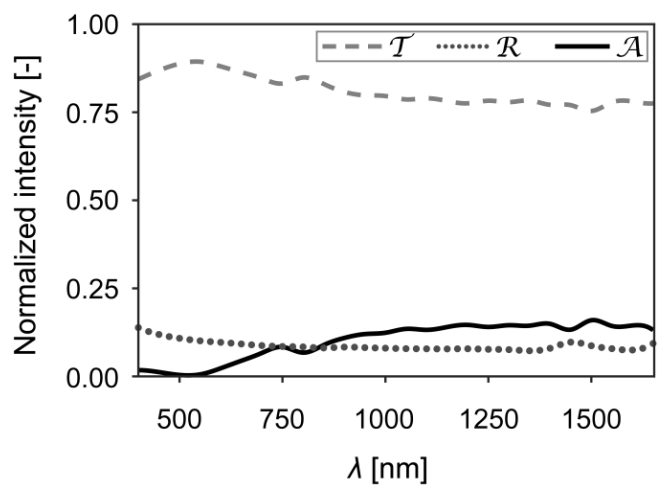

**Figure S5. Spectral properties of the control (glass) sample.** Reflection $(\mathcal{R})$, transmission $(\mathcal{T})$, and absorption $(\mathcal{A})$ spectra of the control sample for visible and near infrared wavelengths (400–1650 nm).

## Supporting Information Videos

**Video S1. Antifogging behavior of metasurfaces.** Comparison of an untreated metasurface to untreated and hydrophobic control glass samples as illustrated in Figure 5. Illumination power density here was 1 sun.